\documentclass[aps,pra,twocolumn,groupedaddress,showpacs]{revtex4-1}

\usepackage{graphicx}
\usepackage[ansinew]{inputenc}
\usepackage{array}
\usepackage{color}
\usepackage{amsmath}
\usepackage{amsxtra}
\usepackage{amstext}
\usepackage{amssymb}
\usepackage{latexsym}
\usepackage{float}

\begin{document}


\title{Attosecond Transient Absorption in Dense Gases: Exploring the Interplay between Resonant Pulse Propagation and Laser-Induced Line Shape Control}

\author{Chen-Ting Liao}
\affiliation{College of Optical Sciences and Department of Physics, University of Arizona, Tucson, Arizona 85721, USA}

\author{Seth Camp}
\affiliation{Department of Physics and Astronomy, Louisiana State University, Baton Rouge, Louisiana 70803, USA}

\author{Kenneth J. Schafer}
\affiliation{Department of Physics and Astronomy, Louisiana State University, Baton Rouge, Louisiana 70803, USA}

\author{Mette B. Gaarde}
\email[]{gaarde@phys.lsu.edu}
\affiliation{Department of Physics and Astronomy, Louisiana State University, Baton Rouge, Louisiana 70803, USA}

\author{Arvinder Sandhu}
\email[]{sandhu@physics.arizona.edu}
\affiliation{College of Optical Sciences and Department of Physics, University of Arizona, Tucson, Arizona 85721, USA}



\begin{abstract}

We investigate the evolution of extreme ultraviolet (XUV) spectral lineshapes in an optically-thick helium gas under near-infrared (IR) perturbation. In our experimental and theoretical work, we systematically vary the IR intensity, time-delay, gas density  and IR polarization parameters to study lineshape modifications induced by collective interactions, in a regime beyond the single atom response of a thin, dilute gas. In both experiment and theory, we find that specific features in the frequency-domain absorption profile, and their evolution with propagation distance, can be attributed to the interplay between resonant attosecond pulse propagation and IR induced phase shifts. Our calculations show that this interplay also manifests itself in the time domain, with the IR pulse influencing the reshaping of the XUV pulse propagating in the resonant medium. 

\end{abstract}



\maketitle


\section{Introduction}
\label{intro}

Important experimental and theoretial advances have been made in the field of transient absorption spectroscopy in the past few years due to the application of novel attosecond ($10^{-18}$ sec) extreme ultraviolet (XUV) light sources\cite{Krausz.2001}. Since 1990's, conventional femtosecond transient absorption spectroscopy has been routinely applied as a powerful tool to study phenomena ranging from basic photophysical and photochemical processes to the workings of biological complexes responsible for vision\cite{Schoenlein.1991.vision}, light-harvesting\cite{Berera.2009.Photosynthesis} {\it etc}. Attosecond light sources, either in the form of attosecond pulse trains (APT) or isolated attosecond pulses\cite{Leone.2008.ATA.Review, Ursula.2012.Review}, have led to a major improvement in the temporal resolution of transient absorption spectroscopy, opening the pathway for direct observation of electron dynamics on their natural timescales. The attosecond approach offers immense potential for the study of complex electronic processes, {\it e.g.} correlation driven charge migration, which occur on attosecond to few-femtosecond timescale\cite{Kuleff2014}. 

Several recent attosecond transient absorption (ATA) experiments have demonstrated the observation and manipulation of electron wavepacket dynamics in dilute atomic gases, such as krypton\cite{Goulielmakis.2010.ATA.Kr.1st, Pabst2012}, argon\cite{Zenghu.2010.ATA.Ar}, neon\cite{Zenghu.2013.Ne.ATA}, helium\cite{Keller.2011.ATA.He, Zenghu.2012.ATA.He.StarkShift, Zenghu.2013.Brought2Light}, dilute molecular gases such as bromine\cite{Leone.2013.ATA.Br2}, and solid-state thin films such as silica\cite{Schultze.2013.ATA.SiO2}, cobalt oxide\cite{Jiang2014.ATA.Coalt}, etc. These efforts have resulted in better understanding of many fundamental physical phenomena including valence electron motion, autoionization, transient electric conduction, light-induced virtual states \cite{Chen.2012.ATA.He.LIS}, gain/loss mechanisms\cite{Keller.2014.ATA.He.GainLoss}, and lineshape modification\cite{Zenghu.2010.ATA.Ar, Pfeifer.2013.LorentzMeetsFano, Pfeifer.2014.ATA.He, Argenti2015}. 

In the majority of ATA experiments and calculations on gaseous samples, however, the temporal and spectral properties have been studied assuming a single-atom response, i.e., the dilute gas assumption in which absorption follows the Beer-Lambert law. Few attempts have been made to observe \cite{Leone.2013.MacroHe, Liao2015} or calculate \cite{Mette.Ken.2012.ATA.He.Reshaping, Chu2013, Mette.Ken.2013.ATA.He.macroscopic3level, Chen.2013.LIP, Leone.2013.MacroHe, Perfetto2015a, Perfetto2015b}  ultrafast transient absorption in dense media, where the collective interaction effects and nonlinear response of the electric dipoles or atomic polarizations are important. Recently, we found experimentally and theoretically that when a dense medium is investigated with ATA, the macroscopic XUV pulse propagation effect cannot be ignored, and the complicated coupling between XUV field and atomic polarizations must be taken into account \cite{Liao2015}. In particular, we observed clear signatures of resonant pulse propagation (RPP) effects in the XUV absorption spectra \cite{Liao2015}. 

\begin{figure}
	\includegraphics[width=6.5cm]{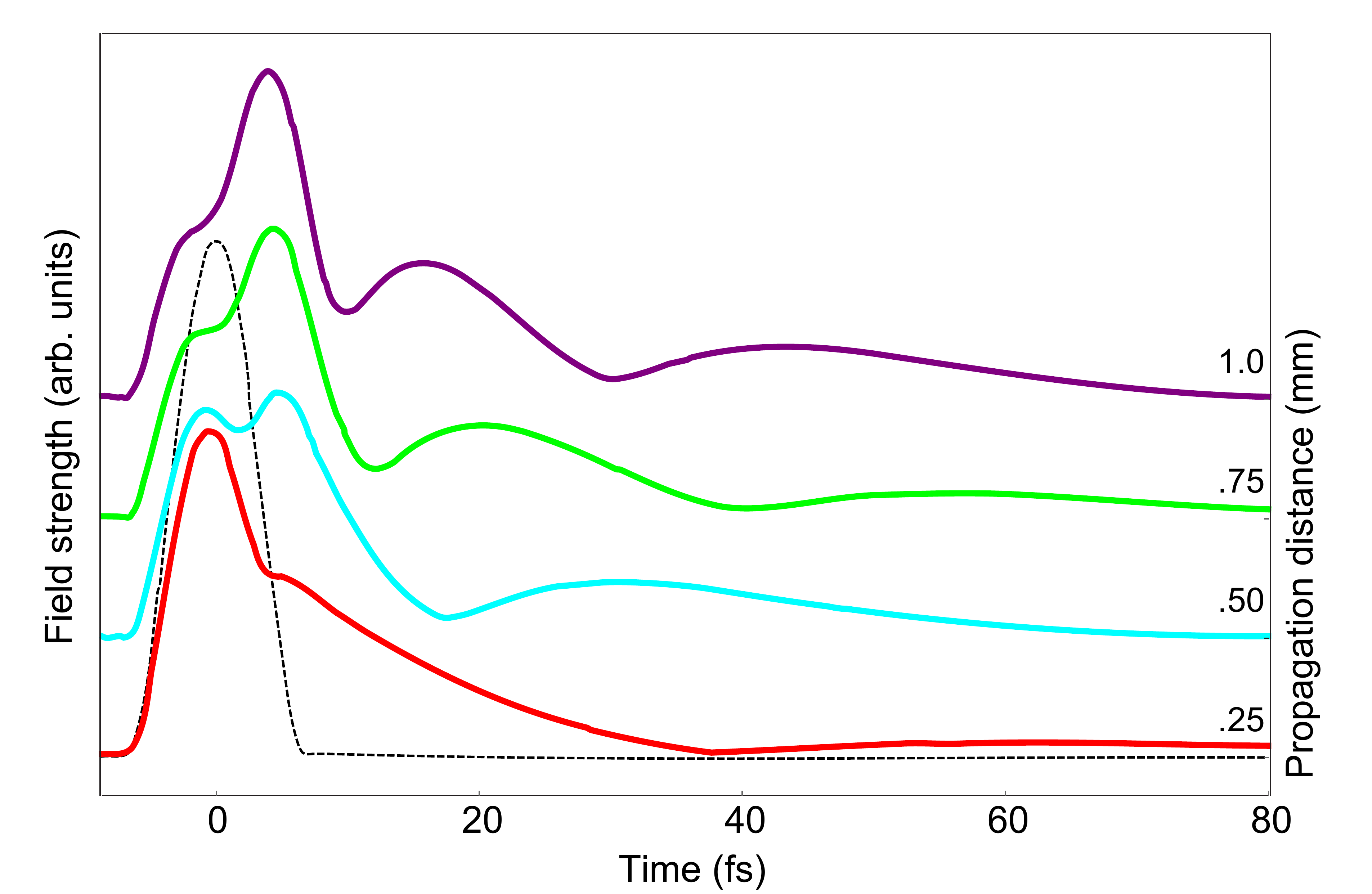}
	\caption{\label{Fig0} Illustration of resonant pulse propagation of a 5 fs XUV pulse propagating through a 16 Torr helium gas in which the XUV light excites a resonance with a 60 fs decay time. We show the XUV electric field strength at different propagation distances up to 1 mm, with the initial pulse shown as the dotted line.  The XUV pulse quickly develops a long tail consisting of a series of sub-pulses which shorten with propagation.
	}
\end{figure}

RPP was first discussed in the 1970s \cite{Crisp1970, GLamb1971} and describes a very general phenomenon in which a short pulse propagating through a medium with  long-lived resonances will undergo strong temporal reshaping. Detailed analysis of the effects of RPP is available for the propagation of visible and near-infrared (IR) femtosecond and picosecond pulses \cite{Bouchene2002, Bouchene2007}, and more generally RPP has been observed and utilized in a range of applications from nuclear magnetic resonance to quantum-well exciton studies \cite{vanBurck.Review}. More recently, it has been discussed in the context of propagation of zero-area, single-photon pulses \cite{Bellini2016}. The temporal reshaping caused by RPP is illustrated in Fig.~\ref{Fig0} for an XUV pulse with a duration $\tau_{xuv}=5$ fs propagating through an 16 Torr helium gas in which it is resonant with a transition with a decay time $\gamma=60$ fs. The figure illustrates how the short pulse initially builds up a long tail which then subsequently develops into a series of sub-pulses. Each sub-pulse is longer in duration and weaker than the previous sub-pulse, and in each sub-pulse the electric field changes its phase by $\pi$ relative to the previous sub-pulse (not shown). The RPP effect is driven by the large spectral phase which can be accumulated due to dispersion (each accumulated phase of $\pi$ in the spectral domain leads to a sign change of the electric field in the time domain) and leads to reduced absorption compared to Beer-Lambert's law \cite{Crisp1970}. In the limit of very long lifetime ($\gamma >> \tau_{xuv}$), the electric field envelope as a function of propagation distance $z$, pressure $P$, and time $t$ will be proportional to the first order Bessel function \cite{Bouchene2002}:
\begin{equation}
{\cal E}(z,t) \propto \frac{J_1(\sqrt{a \sigma_0 \gamma P z t})}{\sqrt{t}},
\end{equation}
where $\sigma_0$ is the absorption cross section on line center and $a$ is a proportionality constant. For a given system the magnitude of the reshaping and the duration of the sub-pulses are thus determined by the pressure-length product $Pz$, and are strongly influenced by the oscillator strength and lifetime of the transition. 

This paper is intended as a systematic experimental and theoretical investigation of the XUV spectral line shapes associated with ATA in a dense helium gas sample that has been laser-dressed by a moderately strong infrared (IR) field of varying intensity and polarization, over a range of time-delays and gas pressures. We are in particular interested in exploring how the interplay between the microscopic IR-laser-dressing and the macroscopic RPP reshaping affects the transient absorption of the attosecond XUV pulse. The RPP effect in our experiment leads to the temporal reshaping of the probing XUV, while the {IR perturbation} facilitates the spectral measurement of this effect because of its role as a time-dependent phase-shift of XUV induced dipole polarization. Apart from exploring a new temporal and spectral regime, our work, unlike earlier RPP studies, employs a two-color IR-pump XUV-probe configuration. Our results elucidate the complex interplay between the XUV and IR pulses, enabled by the nonlinear medium, in both time and frequency domain descriptions. We find that the propagation-induced XUV reshaping influences conclusions and interpretations one can draw from IR-induced spectral line shapes as a function of XUV-IR delay, IR intensity, IR polarization and gas sample density. Reversely, we also find that the IR pulse influences the RPP induced temporal reshaping, especially when the IR pulse is long. We believe that incorporating such macroscopic propagation effects will become more and more important as pump-probe ATA spectroscopy moves towards investigations of complex systems, e.g., dense plasmas, condensed phase systems, and biological samples. 
\begin{figure}
	\includegraphics[width=8cm]{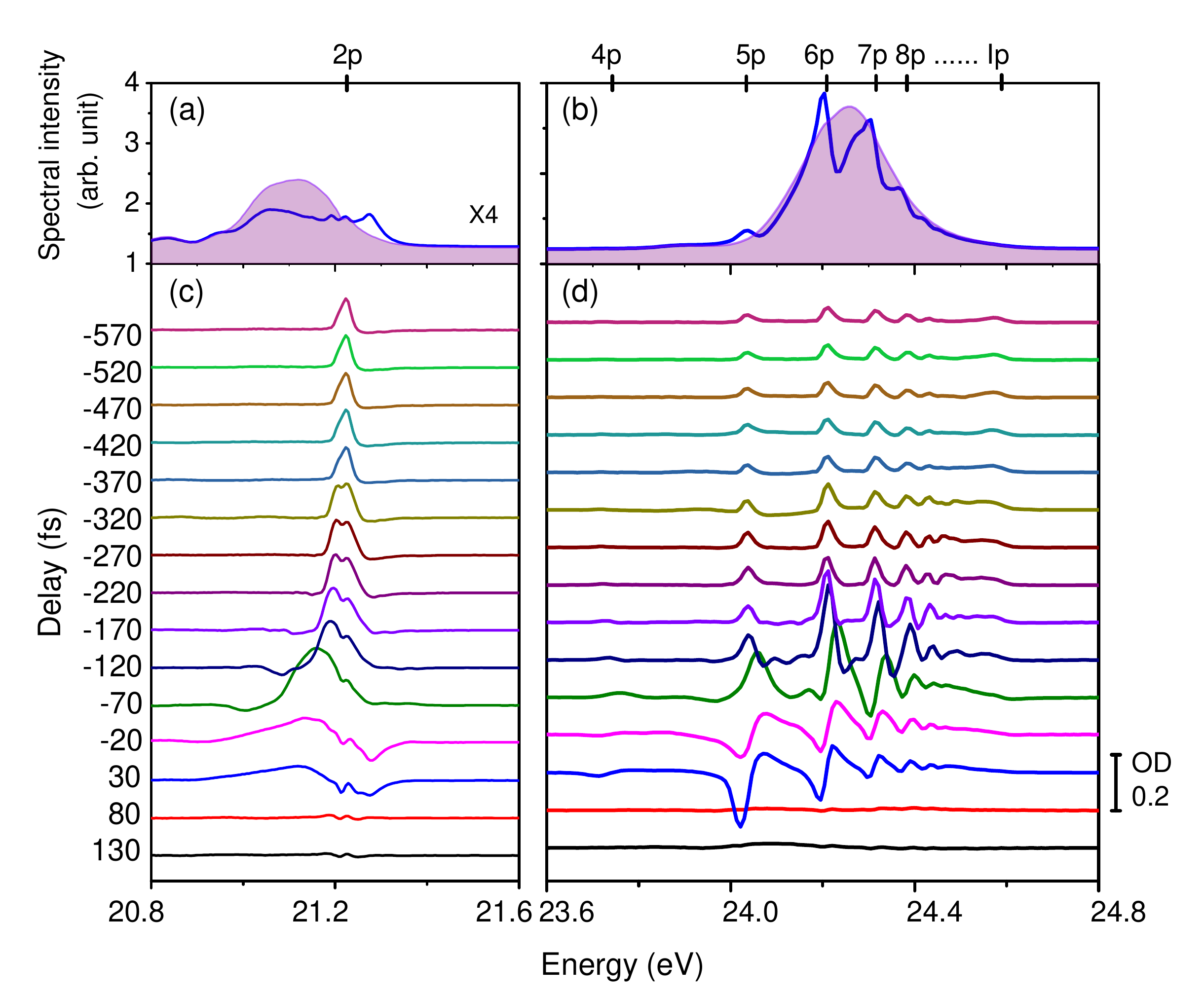}
	\caption{\label{Fig1} Experimentally measured spectral lineshape evolution as a function of delay. (a) and (b) shows transmitted (blue curves) and reference (purple shaded area) XUV spectra around H13 and H15, respectively. (a) is scaled up by a factor of 4 in comparison to (b) for a better view.  (c) and (d) show the evolution of the $OD$ as a function of XUV-IR delay. The scale bar on the right represents signal magnitude corresponding to $OD = 0.2$.
	}
\end{figure}

The paper is organized as follows: We begin by discussing our experimental and theoretical methods in Section \ref{methods}. In Section \ref{evolution} we discuss the ATA spectral line shape in a dense laser-dressed helium gas as it evolves with respect to the relative XUV-IR delay and give a general introduction to the interplay between RPP and the laser-dressing effects. Sections \ref{IRintensity}, \ref{PropagationEffects} and \ref{PolarizationDependent} present experimental and theoretical studies of the line shape dependence on the IR intensity, the gas pressure, and the relative XUV-IR polarization, respectively. We end the paper with a brief conclusion in Section \ref{Summary}.

\section{Experimental and numerical Methods}
\label{methods}
Our experimental setup is discussed in \cite{Liao2015}. Briefly, a Ti:Sapphire laser amplifier produces $\sim$40 fs IR pulses at 1 kHz repetition rate with pulse energy 2 mJ, central wavelength 786 nm, and full width at half maximum (FWHM) bandwidth 26 nm. The carrier envelope phase is not stable and thus averaged in our measurements. The amplified IR pulse is split into two paths. In one path, the IR pulse is focused into a xenon filled hollow-core waveguide (with backing pressure 20 torr and capillary waveguide length 3 cm and inner diameter 150 $\mu$m) to generate an XUV APT via high harmonic generation. The APT is dominated by harmonics 13, 15, 17 (H13, H15, H17), centered at energies 21.03 eV, 24.05 eV, 27.35 eV, respectively, and exhibits individual 440 attosecond bursts and a $\sim$5 fs overall pulse envelope. We monitor the absorption of  H13 and H15 which cause excitations to the $1s2p$ and $1snp$ $(n = 4, 5, ...)$ states of helium. The XUV is combined with the time-delayed collinear IR pulse from the second path using a 45 degree mirror with a hole, and both beams are focused into a 10 mm long gas cell with aluminum foil as gas barrier. A home-made spectrometer, which includes a concave grating (1200 lines/mm, 1 m focal length), a cooled X-ray CCD, and a 200 nm thick aluminum foil to block transmitted IR, is used to measure XUV spectra with a resolution of $\sim$7 meV at 24 eV. Our delay convention is such that negative (positive) delays means that the XUV pulse arrives before (after) the center of the IR pulse. 

The IR-modified absorption spectra are measured at different XUV-IR delays $\tau$ and characterized by the optical density $OD(\omega,\tau) = -\log_{10}(I(\omega,\tau)/I_0)$, where $I(\omega,\tau)$  is the delay-dependent transmitted  XUV spectral intensity in the presence of the IR pulse, and $I_0$ is the reference spectrum. In our experiment, the transmitted XUV  spectrum at large positive delays ($I_0(\omega,\tau > 1000 fs$) is approximately the same as the transmitted XUV spectrum in the absence of the IR pulse, and can therefore be used as the reference spectrum $I_0$ to compute the $OD$ \cite{Leone.2008.ATA.Review}. Each spectrum is retrieved from an average of 11 frames on a X-ray CCD camera, and each frame is acquired with 1 second exposure time, so every measured $OD$ spectrum represents an average of $>$10,000 IR laser shots at each delay, and this is repeated for various IR pulse intensities and polarizations.

Our method for calculating the macroscopic $OD$ spectra is described in detail in \cite{Mette.Ken.2012.ATA.He.Reshaping}. Briefly, we numerically solve the coupled time-dependent Schr{\"o}dinger equation (TDSE), in the single active electron approximation, and the Maxwell wave equation (MWE), for the full two-color IR+XUV field. This yields the delay-dependent spectral intensity of the XUV field at the end of the gas, $I_{out}(\omega, \tau)$, and we then calculate the  $OD$ as described above, using the XUV field at the beginning of the gas as the reference spectrum. We choose parameters for the two pulses and the gas to be close to those of the experiment, to within computational limits. The APT used in the calculations is synthesized from H13-H17 with initial relative strengths of 1:10:6 (this is the same as the experiment), and all initially in phase. The FWHM duration of the APT is $\sim$5 fs, and the peak intensity is $10^{10}$ W/cm$^2$. The IR pulse is centered at 770 nm and has a FWHM duration of 33 fs, unless otherwise specified. We consider a cell of length of 1 mm (not 10 mm), and choose the range of pressures so that the evolution with increasing pressure is similar to that of the experiment. As discussed in the introduction, it is the pressure-length product (usually referred to as the optical thickness) that determines the propagation dynamics and therefore need to be matched between experiment and theory. 

Each calculation shown for a specific delay is averaged over one half cycle, symmetrically around the labeled delay, to mimic the experimental delay resolution and carrier-envelope phase instability. A decay time of  $\sim$60 fs is imposed on the time-dependent atomic polarization, to match the observed (spectrometer resolution limited) absorption linewidths at low gas density, in the presence of the IR \cite{Chen.2012.ATA.He.LIS}. Note that since the pseudo-potential used in the TDSE calculations predicts a bound state energy for the $1s2p$ state of 21.1 eV (instead of the experimentally measured 21.2 eV), the energy range shown in the theory figures is generally shifted slightly relative to the experimental figures. 
 
\begin{figure}[]
	\includegraphics[width=8cm]{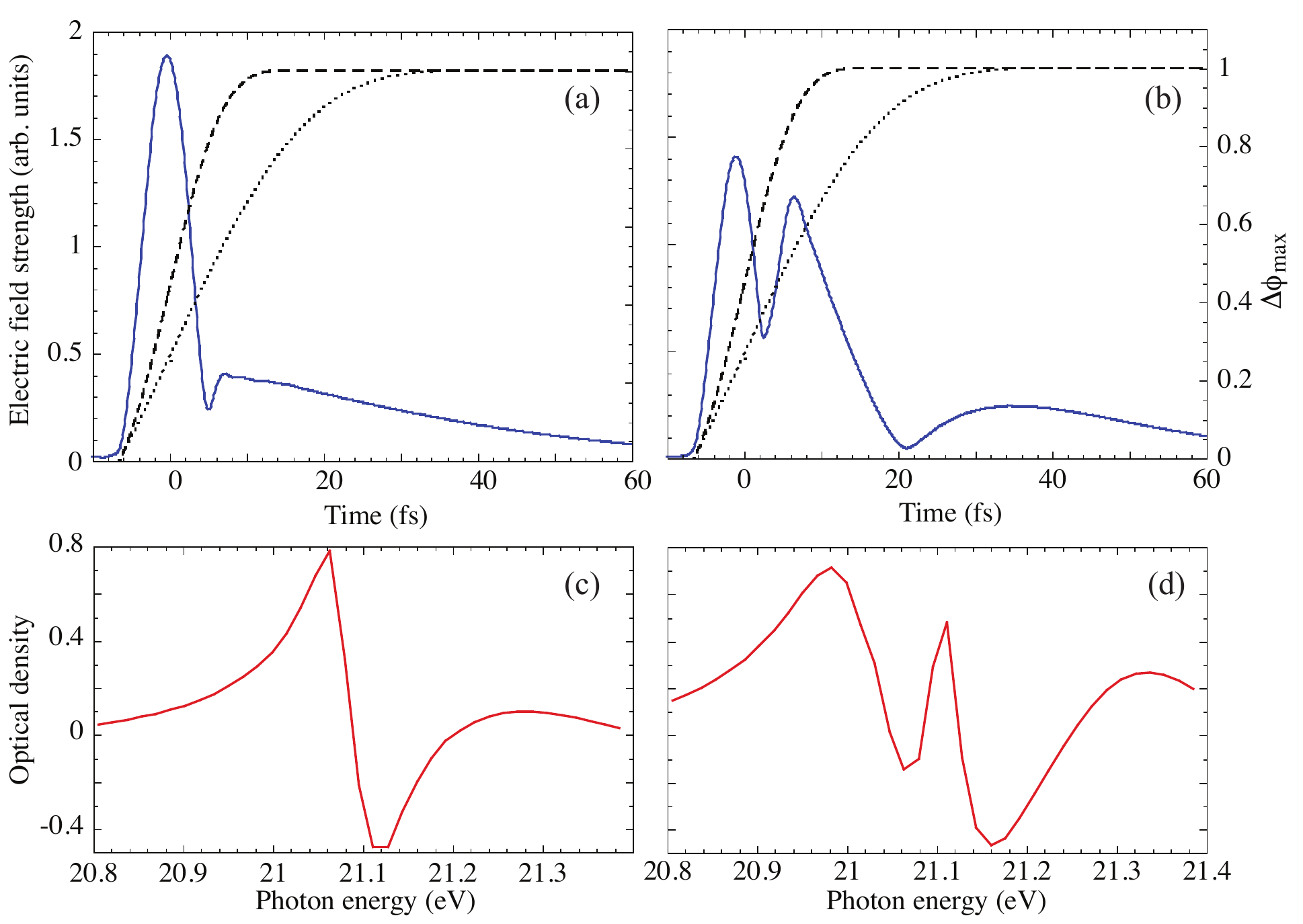}
	\caption{\label{FigModel} (a,b) Calculated XUV time profiles at two different propagation distances $z=0.25$ mm (a) and $z=1$ mm (b) in an 8 Torr helium gas, dressed by a 2.25 TW/cm$^2$ IR pulse, with their corresponding optical density spectra shown above.  In (a) one long sub-pulse is visible in addition to the initial pulse centered on $t=0$, and in (b) two sub-pulses are visible. The figure also shows an approximate time-dependence of the phase shift imposed by two different IR pulses of duration 13 fs (thin dashed lines) and 33 fs (thin dotted line).{(c,d) Spectral lineshape (OD) obtained in the presence of IR pulse for the two propagation distances of 0.25 mm (c) and 1mm (d)}.}
\end{figure}

\begin{figure*}[]
	\includegraphics[width=1 \textwidth]{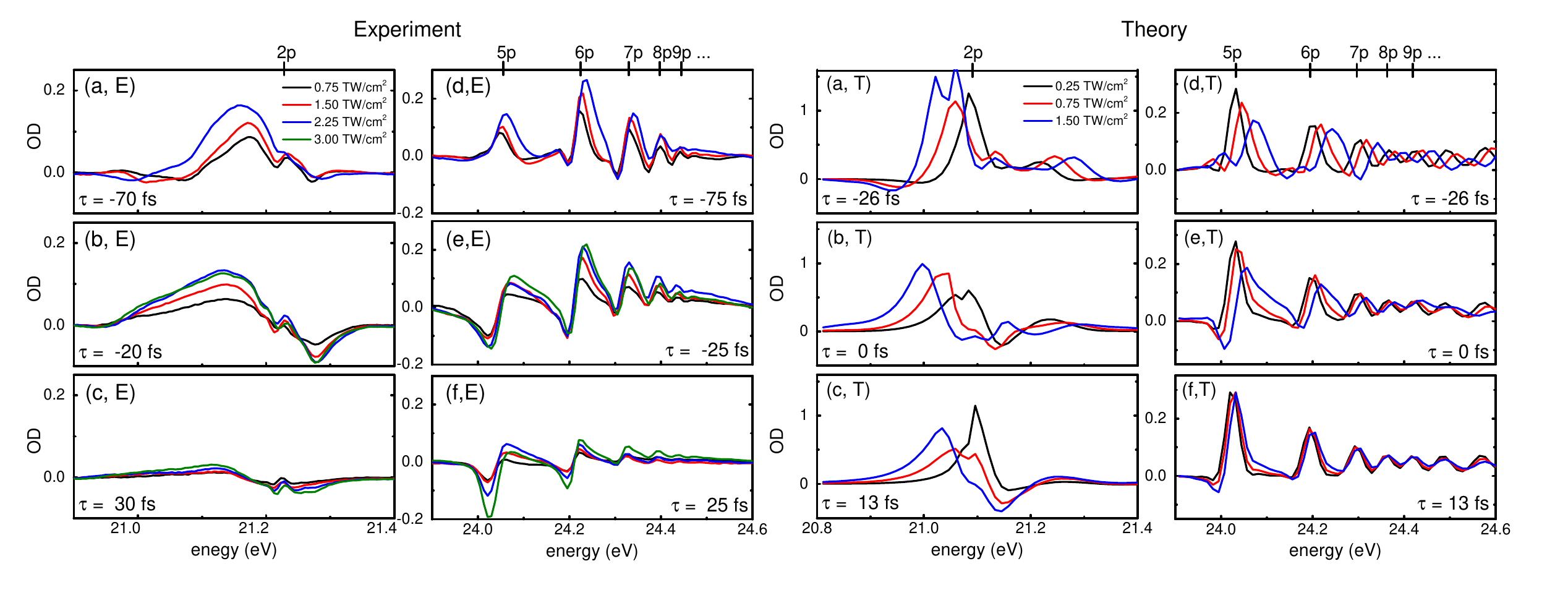}
	\caption{\label{FigIntensity} (a,E) - (f,E): Experimentally measured IR intensity dependent  $OD$ spectra around the $1s2p$ (first column)  and $1snp$ (second column) lines at some representative delays. (a,T) - (f,T): Calculated  $OD$ spectra  at some comparable delays. Note that RPP induced peaks are only visible in the $1s2p$ state.}
\end{figure*}



\section{Spectral lineshape evolution}
\label{evolution}
We start by investigating the evolution of the  $OD$ with respect to the relative delay. Fig. \ref{Fig1}(a) and \ref{Fig1}(b) show the transmitted (blue curves) and reference (purple shaded area) XUV spectra of H13 and H15 at $\tau\sim0$. The helium backing pressure is 8 torr and the IR intensity used is $\sim$3 TW/cm$^2$. The relevant He field-free energy levels, e.g. $1s2p$, $1s4p$, $1s5p$, ..., and up to the ionization potential ($I_p$), are labeled at the top of the figure.

The transmitted spectra illustrate that at some XUV photon energies, the spectral intensity is larger than the reference spectra. Unlike traditional static absorption spectra, the presence of the external IR field modifies the time-dependent dipole moment which has been initiated by the XUV field, predominantly by the addition of a time-dependent phase due to the AC Stark shift of the excited state \cite{Zenghu.2010.ATA.Ar, Pfeifer.2013.LorentzMeetsFano, Chen.2013.LIP}. In the spectral domain, this leads to a redistribution of energy between the light field and the atom and generally gives rise to transient absorption profiles that include both positive and negative values, corresponding to absorption (loss) or  emission (gain) at different  frequencies. In the simplest possible case in which one starts with a purely Lorentzian absorption line shape around a bound state resonance and then adds an instantaneous phase shift due to a short IR pulse, the resulting absorption line shape can be varied smoothly between a Lorentzian, a dispersive (Fano, \cite{Fano1961}), and a window resonance line shape depending on the magnitude of the {laser induced phase (LIP)} shift. This was discussed extensively in \cite{Pfeifer.2013.LorentzMeetsFano} and other papers. 

Fig. \ref{Fig1}(c) and (d) show the $OD$ spectral lineshape evolution as the delay changes. For large positive delays, $\tau$ $>$ {70} fs, all of the IR pulse arrives before the XUV pulse and the  $OD$ is flat $\sim 0$ which means that there is no change in the absorption relative to the reference spectrum. Since the IR pulse is too weak to excite the helium atoms in the absence of the XUV field, at these delays the only absorption is due to the unperturbed bound states which cannot be resolved by our spectrometer due to their long lifetimes. 
In contrast, at large negative delay, e.g., $\tau$ $<$ -370 fs, when the XUV comes first, we observe traditional near-Lorentzian (Voigt) line shapes in the  $OD$ spectra. The Voigt profile results from the convolution of two broadening mechanisms manifesting as Lorentzian and Gaussian, where the Gaussian part is due to Doppler broadening, pressure broadening, and the profile of the IR pulse. The Voigt lineshapes at large negative delays are due to the perturbed free induction decay of the excited populations in $1s2p$ or higher lying $1snp$ states. Upon the arrival of the XUV pulse, H13 and H15 coherently prepare two atomic polarizations corresponding to  the $1s2p$ and $1snp$ states. These atomic polarizations freely evolve till the IR arrives at some time later, and perturbs the typical free induction decay process. Compared to the natural lifetime of the $1snp$ states ($\sim$nanosecond range), the femtosecond IR  perturbation acts like an impulsive kick, giving rise to a Voigt lineshape whose width is primarily a characteristic of the IR pulse duration. We find that this constant width Voigt lineshape persists in the $OD$ even for  very large negative delays (more than a few ps, not shown) in our experiments. 

At delays close to zero, complicated line shapes can be observed in the $OD$ spectra. In the intensity regime shown in the figure, the overall line is shifted to lower energy and the shape is Fano-like, although more complex, with several sub-features close to line-center and additional features in the wings of the profile. The downward Stark shift is caused by the laser-induced interaction with higher-lying excited states ({see for instance} \cite{Chen.2013.LIP,Niranjan.2012.VMI.He}). The overall dispersive-like shape is due to the laser-induced phase (LIP) discussed above. The features in the wings of the profile result from the relatively long duration of the IR pulse, as the LIP perturbation is imposed not instantaneously but over a finite time. Finally, the narrow features close to the field-free $1s2p$ energy are due to RPP \cite{Crisp1970, GLamb1971} as we discussed in  \cite{Liao2015}. In the following, we will illustrate the interplay between the RPP XUV temporal reshaping and the effect of the IR pulse in further detail. 


This interplay is illustrated in Fig. \ref{FigModel} which shows calculated time profiles of the XUV electric field strength and the corresponding optical densities at two different propagation distances in an 8 Torr He gas. The IR pulse is short, with a FWHM pulse duration of 13 fs and a peak intensity of 2.25 TW/cm$^2$. Since the biggest dipole response is from the $1s2p$ transition which is excited by H13, we show time profiles corresponding to the H13 part of the APT. Since the RPP effect is driven by the spectral phase accumulated via dispersion by the propagating, resonant XUV pulse it is to first order independent of the IR pulse, and similarly to Fig. 1 one can observe the build-up of the long tail in the XUV time profile and the formation of multiple sub-pulses as the pulse propagates \footnote{The depth of the first minimum, between the main pulse and the first sub-pulse is influenced by the IR and for instance depends on the sub-cycle delay. This short-time modulation is due to the excitation of light-induced states in the vicinity of the $1s2p$ state which are within the bandwidth of the XUV pulse and which can be excited while the IR field is on, see for instance \cite{Chen.2012.ATA.He.LIS}. }. We note that the duration of the first sub-pulse acts as a new effective lifetime for the resonant interaction. The shortening of the first sub-pulse during RPP therefore in general leads to broadening of the absorption profile, resulting in a width proportional to the optical thickness $Pz$ \cite{vanBurck.Review}. 


Fig. \ref{FigModel} also illustrates the effect of the IR pulse by showing an approximation to the LIP of the time-dependent dipole moment, assuming that the phase accumulates proportionally to the IR intensity. When the IR pulse is short the phase shift is imposed almost instantaneously compared to the lifetime of the dipole moment. For the parameters shown in Fig.~\ref{FigModel}, the total phase shift $\Delta\phi_{max}$ is approximately $\pi/2$, giving rise to a mostly dispersive shape of the absorption profile as shown in the {Fig. \ref{FigModel}(c) and (d)}. As the XUV pulse propagates, the overall absorption profile broadens  due to the reduced effective lifetime as describe above, but its shape does not change substantially since this is controlled by the size of the LIP. However, when the XUV pulse has propagated enough that the second sub-pulse appears in the XUV tail (as in Fig.~\ref{FigModel}(b)) a new narrow feature appears of line-center in the absorption profile (Fig.~\ref{FigModel}(d)). This is because the second sub-pulse is phase-shifted by $\pi$ relative to to the first sub-pulse and therefore generally will give rise to absorption (as opposed to the emission on line center caused by the LIP). 
Additional narrow features would appear in the absorption spectrum, nested around line center, for longer propagation distances as additional sub-pulses would appear in the XUV temporal profile. 
For our experimental parameters, we generally observe one or two RPP features close to the $1s2p$ resonance. We do not observe strong RPP effects around the higher lying $1snp$ states as these transitions are weaker and the temporal reshaping is therefore very small. 

Finally, Fig. \ref{FigModel}(a) and (b) also illustrate the time-dependent phase shift that would be imposed by a longer IR pulse, such as that used in the experiment and in the calculations shown in the remainder of the paper. It is clear that the temporally extended phase shift means that at long propagation distances, the full phase shift has not yet been imposed by the time that the first sub-pulse ends. In Section~\ref{PropagationEffects} we will discuss how this affects both the evolution of the spectral absorption profile and the temporal XUV profile during propagation.

\begin{figure}
	\includegraphics[width=7.5cm]{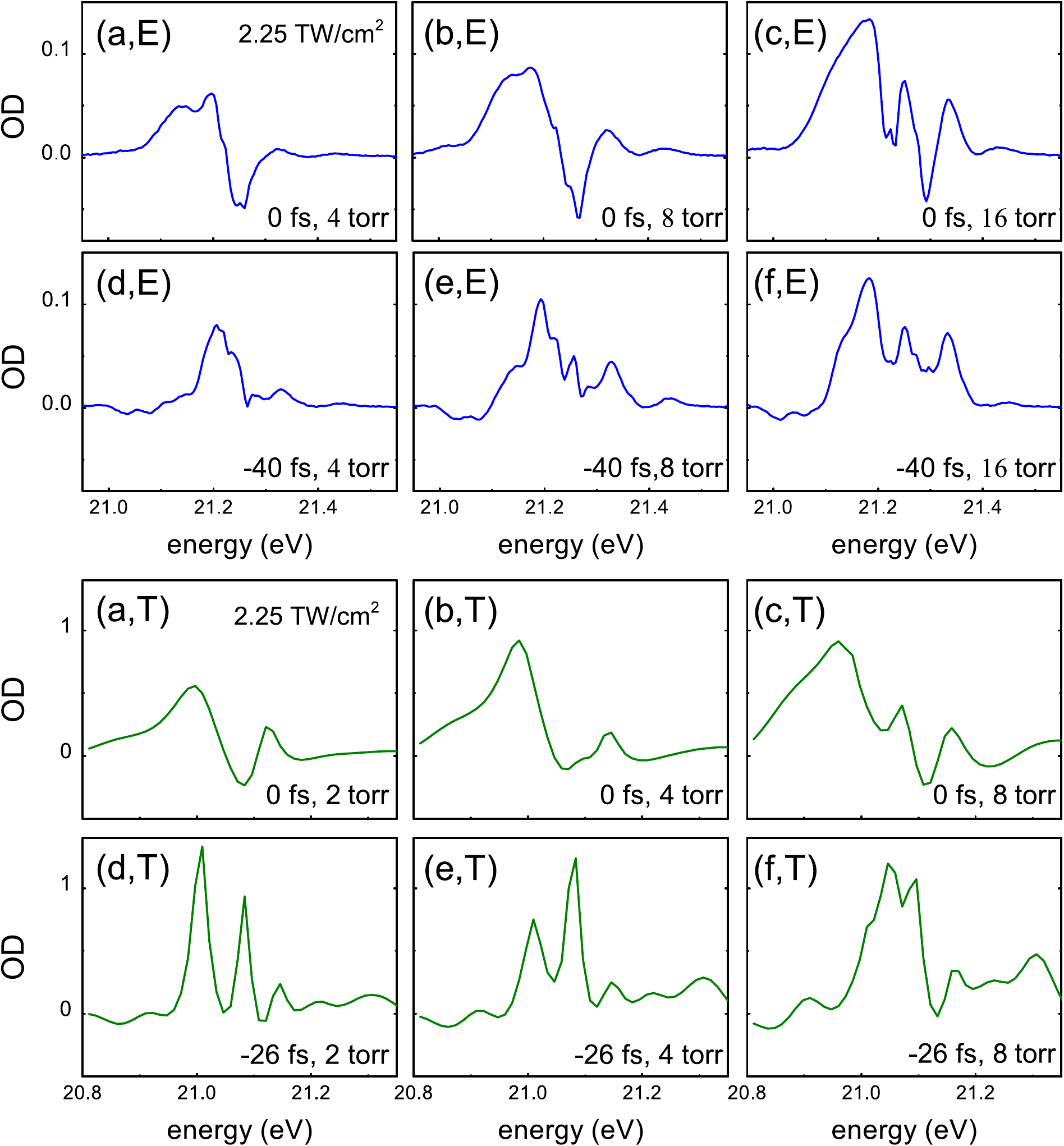}
	\caption{\label{Fig.Pressure}(a,E) - (f,E): Experimentally measured $OD$ spectra around the $1s2p$ line for different pressures and delays. (a,T) - (f,T): Calculated  $OD$ spectra using comparable pressures and delays.}
\end{figure}


\section{IR intensity dependent lineshape modification}
\label{IRintensity}

In this section, we experimentally and theoretically explore the dependence of the absorption line shapes  on the perturbing IR  intensity.  Fig.~\ref{FigIntensity} shows the IR intensity-dependent $OD$ spectra at some representative time delays for both experiment and theory. From top to bottom, the three delays represent a small negative delay, a delay close to overlap, and a small positive delay. For the small negative and the small positive delay, there is limited overlap between the IR and the XUV pulses. Note however that positive and negative delays are not symmetric in the effects of the IR as we discussed above, since the XUV absorption is only altered by the part of the IR pulse that arrives after the XUV excitation. The dressing IR peak intensity in the experiment is estimated to range from 0.75 ($\pm$ 0.5) to 3 ($\pm$ 0.5) TW/cm$^2$, however in the calculations we get the best agreement with the measured line shape at 3 ($\pm$ 0.5) TW/cm$^2$ when we use an IR intensity of 1.5 TW/cm$^2$. For the calculated results, we also show two lower intensities of 0.25 TW/cm$^2$ and 0.75 TW/cm$^2$. We note that the experimental intensity values are overestimated as the loss at the entrance aperture of the gas cell is not taken into account. We should also point out that while the He backing pressure is 8 Torr in the experiment, the actual pressure in the 10 mm long He gas cell is smaller. Furthermore, the gas pressure in the interaction region is not uniform and varies with the size of the laser-drilled entry and exit holes on the gas cell. In all the calculations shown in Fig.~\ref{FigIntensity} we have used a pressure of 4 Torr in the 1 mm He gas cell. This pressure has been chosen to approximately match the size of the RPP peak in the experiment, see also discussion in connection with Fig.~\ref{Fig.ODvsz12}. Lastly, the absolute values of experimental and theoretical OD differ due to the substantial IR background present on our detector, which artificially reduces the experimental OD values obtained from the weak H13 by a large amount, while having minimal effect on OD obtained from H15.

As  can be observed from Fig.~\ref{FigIntensity}, a stronger IR pulse leads to a broadening and a change in the absorption line shape for all of the delays shown in the figure, in both experiment and theory. This agrees with the discussion of the LIP in the previous section - a larger IR intensity leads to a stronger coupling between excited states and therefore a larger Stark shift and a larger LIP. In the theory results, we can follow the evolution of the line shape with intensity from near-Lorentzian to near-dispersive at some delays (see for instance Fig.~\ref{FigIntensity} (c,T) and (e,T)). In the experimental results, we are not able to lower the IR intensity enough to observe near-Lorentzian profiles for the $\tau = -20$ fs and $\tau = +30$ fs cases ((c,E)-(f,E)). This is likely due to a small leakage of the XUV-generating IR beam through the interaction chamber - this leads to a background of IR which is always present around zero delay. We nevertheless do observe increasingly asymmetric profiles as the IR intensity is increased. We note that the RPP peaks in the $1s2p$ line are visible at all intensities and for all delays, in both theory and experiment. This is as expected since the RPP reshaping is driven by the mismatch of the short XUV duration and the long lifetime of the excited states and to first order does not depend on the IR pulse. On the other hand, as we discussed in \cite{Liao2015}, the RPP does have an influence on how the IR-perturbation reveals itself in the XUV absorption spectrum. In the following section, we will explore in more detail how the RPP and the {LIP} interact in our APT transient absorption scenario. 

\section{Gas pressure dependent lineshape modification}
\label{PropagationEffects}

\begin{figure}
	\includegraphics[width=8.5cm]{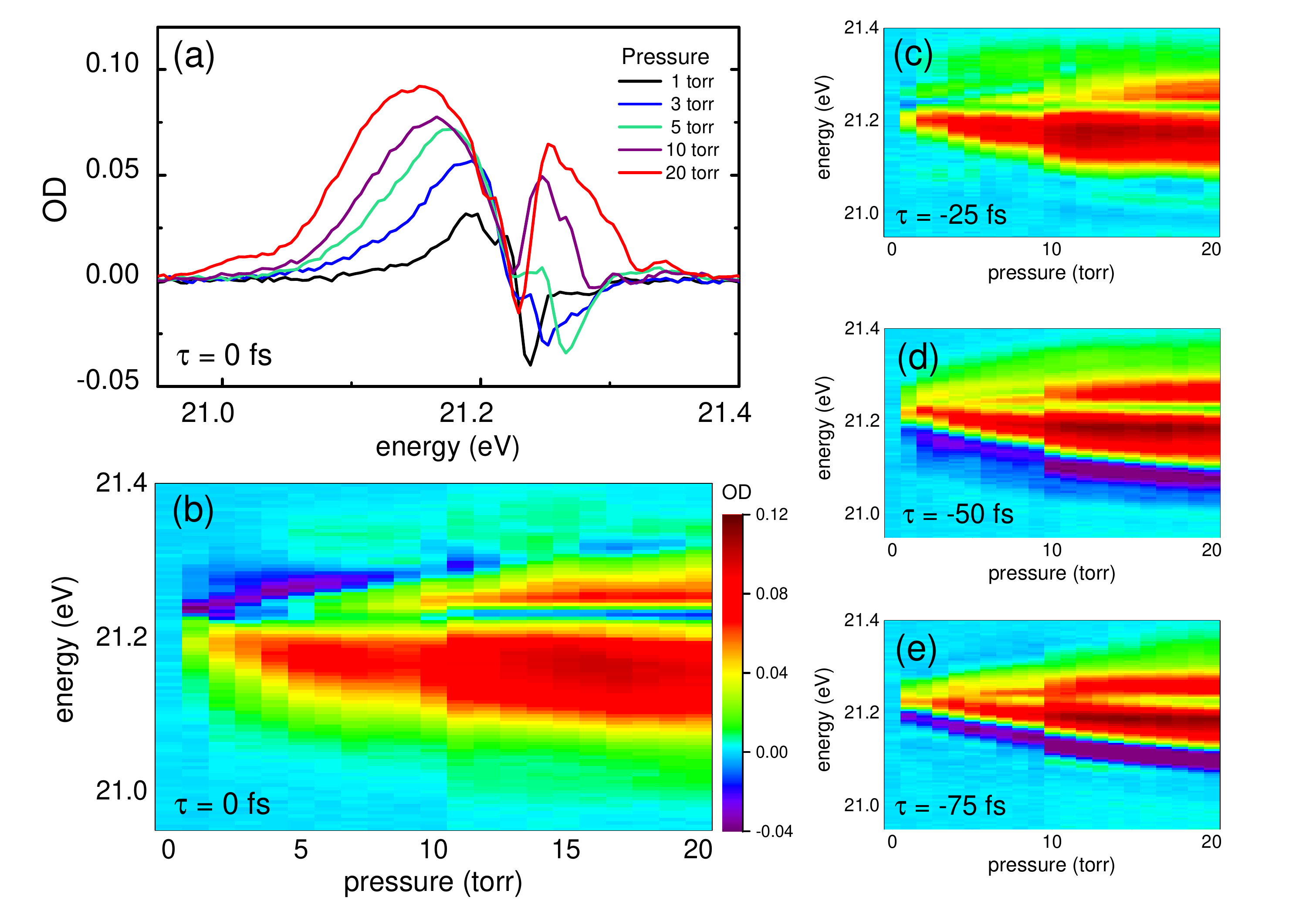}
	\caption{\label{Fig.PressExp} Experimentally measured OD as a function of backing pressure for the case of IR intensity of 2.25 TW/cm$^2$. (a) Shows line-outs of the OD at some representative backing pressures at delay zero (increasing and broadening from smallest to largest pressure), and (b)-(e) shows the full two dimensional map pressure dependence the OD at four representative delays, 0, -25, -50, -75 fs, respectively.}
\end{figure}

\begin{figure}
	\includegraphics[width=8cm]{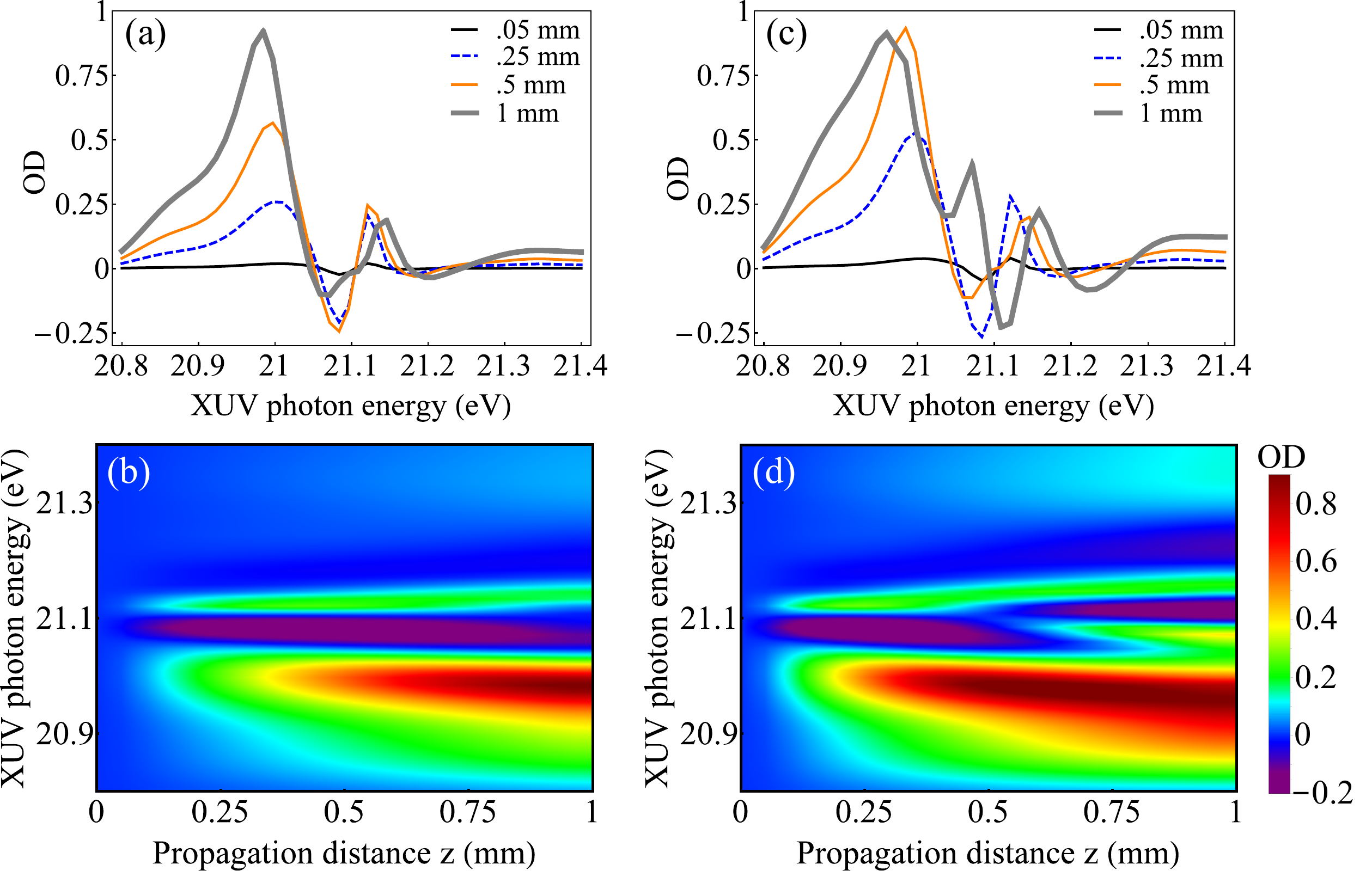}
	\caption{\label{Fig.ODvsz12} Calculated $OD$ as a function of propagation distance for the case of 4 Torr and an IR intensity of 2.25 TW/cm$^2$. (a) Shows line-outs of the  $OD$ at four different propagation distances (increasing and broadening from smallest to largest $z$), and (b) shows the full $z$-evolution of the  $OD$. Note that in (b) the color scale is chosen so that both positive and negative values of the  $OD$D are saturated at the peaks. (c) Same as Fig. \ref{Fig.ODvsz12}(a) except for 8 Torr. (d) Same as Fig. \ref{Fig.ODvsz12}(b) except for 8 Torr.}

\end{figure}

Figs. \ref{Fig.Pressure}, \ref{Fig.PressExp}, and \ref{Fig.ODvsz12} explore the relationship between the RPP-induced and the {LIP based} reshaping of the OD line shape around the $1s2p$ line. In Fig. \ref{Fig.Pressure} we show measured and calculated spectra at three different pressures and two delays. The three pressures are chosen such that at delay zero, the RPP peak is not visible at the lowest pressure (a, E\&T), barely visible at the intermediate pressure (b, E\&T), and prominent at the highest pressure (c,E\&T) in both the experimental and theoretical results. At small negative delay (d,E)-(f,E) and (d,T)-(f,T), we see a similar trend, except the lineshapes generally consist of a narrower main peak and one or several sidebands. 


Fig. \ref{Fig.PressExp} shows a detailed experimental exploration of the pressure dependence of the line shape. Fig. \ref{Fig.PressExp}(a) shows the OD line shape at different pressures, at zero delay, and Fig.\ref{Fig.PressExp}(b)-(e) show full two dimensional maps of the OD vs photon energy and pressure for four different delays. A number of different observations can be made from these figures that all point to the complex nature of the interplay between the macroscopic RPP effect and the microscopic LIP effect: (i) The threshold pressure at which the RPP peak first appears in the OD spectrum is approximately 5 Torr and is independent of the delay. This is in agreement with the observation {from Fig.~\ref{FigIntensity}} in which we  saw that the appearance of the RPP peak was independent of the IR intensity. (ii) The OD line shape is changing with pressure, at most delays starting from a primarily dispersive shape at low pressure to a shape in which the RPP peak on the high energy side is so dominant that there is no longer a minimum - {\it i.e.} the initially dispersive line shape is no longer recognizable. (iii) At delays close to zero (including -25 fs), the bandwidth of the OD profile does increase with pressure, but slower than the linear increase with $P$ one would expect from  RPP \cite{vanBurck.Review}. 

The evolution of the OD spectrum with pressure is equivalent to the evolution of the OD with propagation distance. This is demonstrated in Fig. \ref{Fig.ODvsz12} which shows a theoretical study of the evolution of the OD with propagation distance for two different pressures, 4 Torr (a-b) and 8 Torr (c-d). In the 4 Torr case, the  RPP peak is only just visible at the end of the medium, as a slight increase of the $OD$ on line center (see also Fig.~\ref{Fig.Pressure}(b,T)). The 8 Torr calculation (c-d) shows that, as one would expect, the line shape that has been reached at $z=1$ mm in the 4 Torr case, is reached at $z=0.5$ mm in the 8 Torr case, and then develops further in the remainder of the medium. We can thus directly compare the measured pressure dependences in Fig.~\ref{Fig.PressExp} to the calculated propagation distance dependences in Fig.~\ref{Fig.ODvsz12}.

The calculated results in Fig.~\ref{Fig.ODvsz12} agree with the measured results in Fig. \ref{Fig.PressExp} in terms of points (i)-(iii) made above. We can explain these three observations in the following way, combining our understanding of RPP and the LIP: (i) The threshold pressure represents the pressure at which the XUV pulse develops a second sub-pulse in the tail. This is predominantly controlled by the strength of the ground-to-excited state coupling and the (effective) lifetime of the excited state and therefore to first order independent of the laser parameters. (ii) As the first sub-pulse gets shorter during propagation (and becomes comparable to the IR pulse duration), the propagation of the XUV pulse becomes quite complicated because the time-dependent dipole moment is driven by the combination of an XUV pulse with a long tail and an IR pulse which imposes a phase which varies in time over the entire effective lifetime of the XUV-induced dipole moment. In addition, when the effective lifetime becomes shorter than the IR pulse duration, the full LIP will not be imposed on the dipole moment during its effective lifetime. These effects causes the line shape to dynamically change during the propagation and is the most direct effect of the RPP-LIP interaction. In the calculations (see for instance Fig.~\ref{Fig.ODvsz12}(c)), we observe that the absorption line shape indeed reflects the reduced LIP as the pulse propagates further, changing from closer-to-dispersive to closer-to-Lorentzian.  (iii) The spectral broadening in the calculations also initially quickly increases and then almost saturates at longer propagation distances. This is also an effect of the long IR pulse duration - in \cite{Liao2015} we showed that the RPP spectral broadening with propagation distance is much faster when the laser pulse is short.

\begin{figure}[]
	\includegraphics[width=8cm]{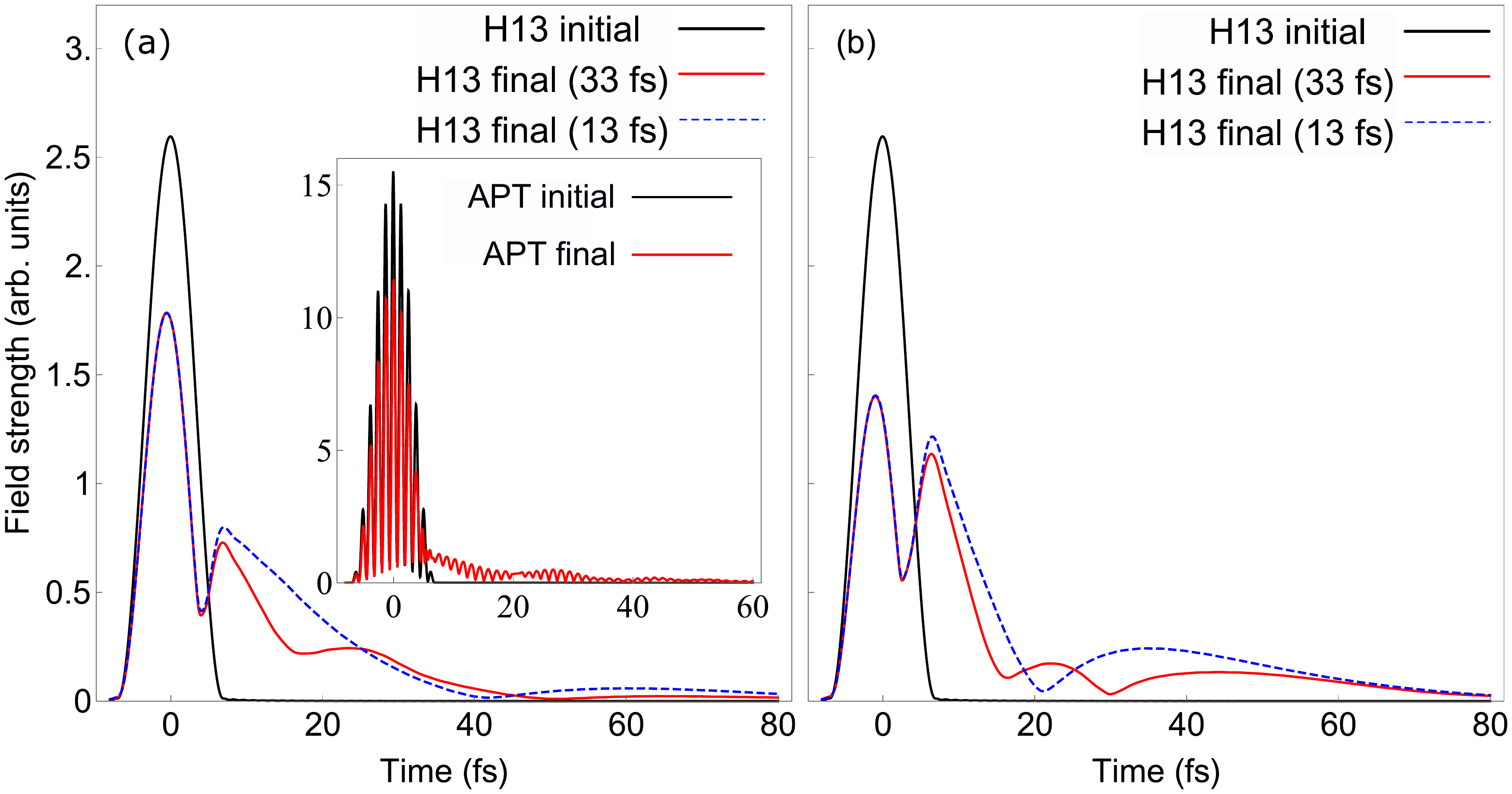}
	\caption{\label{FigTime} Calculated XUV time profiles, at the beginning (black) and end (solid red or dashed blue) of the helium gas at a pressure of 4 Torr (a) or 8 Torr (b), dressed by 2.25 TW/cm$^2$ IR pulse. The final time profiles for two different IR pulse durations are shown, 33 fs (solid red) and 13 fs (dashed blue). The inset in (a) shows the initial and final time profiles of the full APT for the 4 Torr case. }
\end{figure}

The time-domain consequences of the RPP-LIP interplay are illustrated in Fig.~\ref{FigTime} in which we show the initial and final XUV time profiles of the H13 part of the APT at the end of {1mm propagation in He gas at  4 Torr (a) or 8 Torr (b) pressure}, when the gas is dressed by either a 33 fs (solid red) or 13 fs (dotted blue) IR pulse. First, we note that even at 4 Torr, the propagation induced reshaping is very substantial, with the area of the first sub-pulse (between approximately 4 and 40 fs) almost comparable to the area of the main pulse (between -4 and 4 fs). A second sub-pulse is barely visible beyond 40 fs. The appearance of the second sub-pulse at this pressure is consistent with the OD shown in Fig.~\ref{Fig.Pressure}(b,T) and Fig.~\ref{Fig.ODvsz12}(a,b) in which the RPP peak becomes visible at the end of the 4 Torr medium. Second, the additional modulation that can be observed in the middle of the first sub-pulse in the 33 fs IR case results from the interaction with the long IR pulse, since it is absent when the IR pulse is shorter. The RPP-{LIP} interaction is even more clearly manifested in the time profiles at the end of the 8 Torr medium (b). In this case the longer IR pulse causes not only an additional modulation of the first sub-pulse but also differs in the timing of the second sub-pulse by about 10 fs, with the minimum before the second sub-pulse changing from about 30 fs to about 20 fs. We find that if we use shorter and shorter IR pulses, the position of this minimum converges to its IR-free position at about 18 fs. These results indicate the interaction between RPP and the IR-induced control of the time-dependent dipole moment works in both directions - RPP influences the {LIP}-controlled absorption line shape through the appearance of narrow absorption features, and the LIP influences the RPP-controlled temporal reshaping of the propagating pulse in the limit when the time scale of the LIP and the first sub-pulse are comparable. 

The inset in Fig. \ref{FigTime}(a) shows the time profile of the full APT at the beginning and end of the 4 Torr gas jet. Because H13 is only a small fraction of the APT, the tail of the APT looks much smaller compared to the main pulse, but its shape can be recognized from Fig. \ref{FigTime}(a). It is interesting to note that the influence of the dipole response from the $1snp$ states, although weaker than the $1s2p$ response, is visible in the tail in the form of the half-cycle modulation of the tail which would not happen in the absence of radiation around H15\footnote{We note that in this plot we have not averaged over delays as we have in all other plots. We change the delay by shifting the XUV pulse in time and in the delay average the sub-cycle features are therefore strongly suppressed.}.

\section{IR Polarization Dependent Lineshape}
\label{PolarizationDependent}

Finally, we study the influence of the relative polarization of the XUV and IR fields, especially as it relates to the appearance of the RPP peaks in the $1s2p$ absorption profile. All of the results discussed above, both theory and experiment, have been performed using parallel polarizations of the XUV and IR electric fields. We are only aware of one study published so far which studies the influence of the relative XUV and IR polarizations in attosecond transient absorption \cite{Reduzzi2015}. Reduzzi and collaborators studied the polarization-dependent transient absorption of an isolated attosecond pulse centered around the bound states in helium, in the single atom limit. They found that perpendicularly polarized fields lead to a reduced coupling between the XUV-excited manifold of $1snp$ states and nearby dark $1sns$ and $1snd$ states compared to parallel polarizations, and attributed this to the symmetries of the excited states with perpendicular (only $m=1$ states excited) vs parallel (only $m=0$ states excited) polarizations. This means that for perpendicular polarizations, the IR field can only couple excited $1snp$ states to $1snd$ states.

\begin{figure}
	\includegraphics[width=8cm]{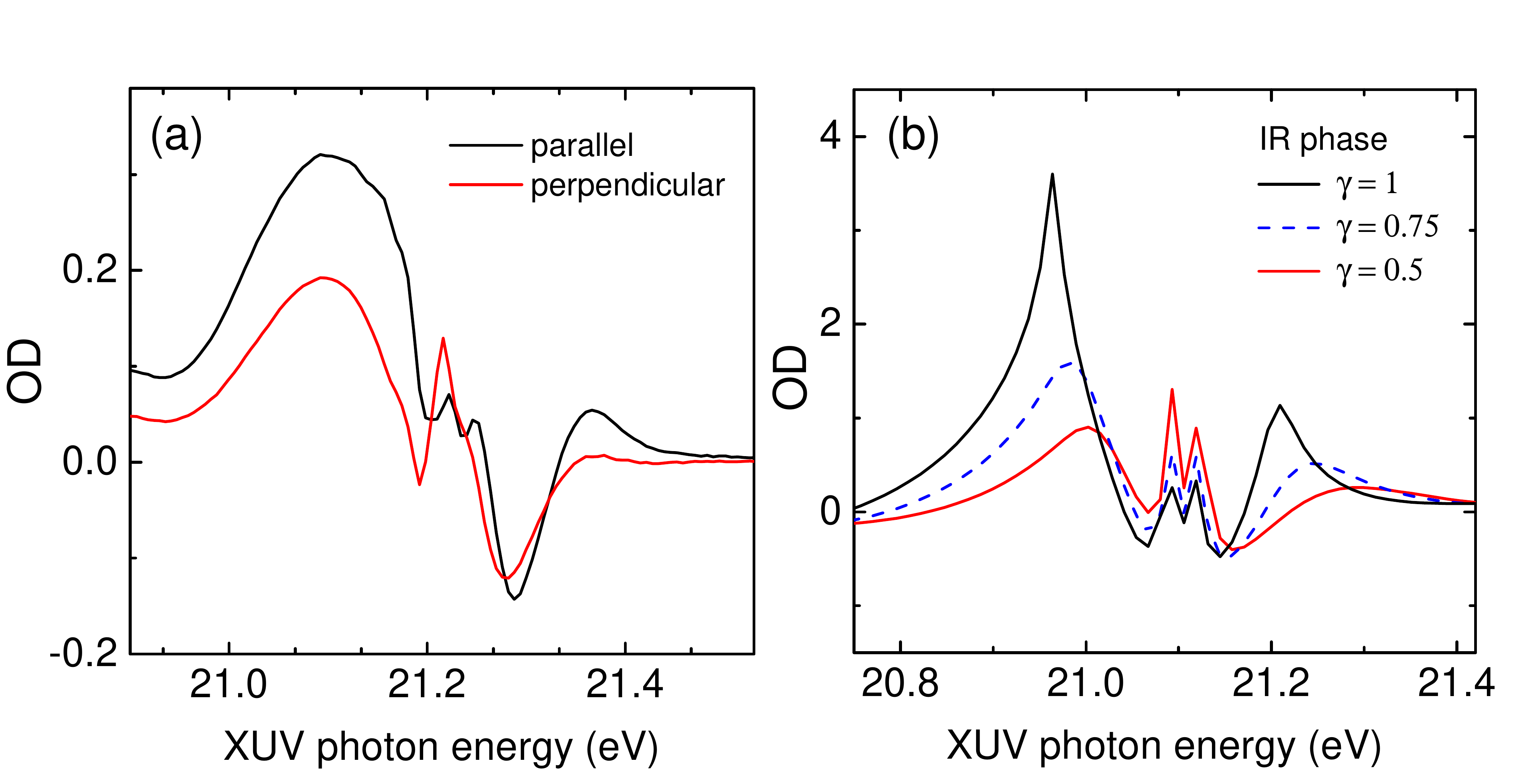}
	\caption{\label{FigPolarization} (a) Experimentally measured spectral lineshape at delays close to zero with different pump-probe polarizations. The black curve is parallel ($0^{\circ}$) geometry and the red one is perpendicular ($90^{\circ}$) geometry. (b) Theory results calculated using the 2-level TDSE-MWE model (see text). We show three different strengths of the IR-induced coupling, modeling the change from parallel polarizations ($\gamma=1$, solid black) to non-parallel polarizations ($\gamma=0.75$, dashed blue, and $\gamma=0.5$, solid red).}
\end{figure}

Fig.~\ref{FigPolarization}(a) compares the {experimental} $OD$ around the $1s2p$ line for parallel (black) and perpendicular (red) XUV and IR polarizations. In both cases, the IR intensity is 1.8 ($\pm$ 0.5) TW/cm$^2$ and the backing pressure is 8 torr. The figure shows that the perpendicular case exhibits a weaker $OD$ overall, but a stronger RPP peak. To mimic this IR-field-polarization dependence theoretically, we have performed a set of model calculations as described in \cite{Liao2015}. Briefly, we solve the MWE coupled to a simpler solution {of} the TDSE for a two-level system interacting with a resonant light field. 
The effect of the IR pulse is modeled as a time-dependent phase that accumulates on the upper state amplitude, proportional to the IR intensity, as if the {LIP} was strictly proportional to the Stark shift. The IR perturbation and the attosecond pulse begin at the same time. A change in the relative XUV and IR polarizations away from parallel is modeled as decrease in the laser-induced coupling (and thereby the LIP), due to the smaller number of states that the $1s2p$ excited state can be IR-coupled to \cite{Reduzzi2015}. In {our calculations}, the XUV pulse is an isolated 400 attosecond XUV pulse centered on 21.1 eV, the IR pulse has a FWHM duration of 13 fs, and we impose a $\sim$ 110 fs decay time on the time-dependent atomic polarization. In Fig.~\ref{FigPolarization}(b) we show results for three different coupling strengths $\gamma$, where the LIP at $\gamma=1$ has been chosen such that the $OD$ line shape and the appearance of the RPP peaks are similar to those of the experiment for parallel polarizations \footnote{We note that the model calculations in general predict $OD$ values much larger than those of the experiment and the full calculations, even when the line shape and the general evolution with propagation distance generally agree. In addition, the experimental peak values of the $OD$ around the $1s2p$ line are artificially reduced by the background of IR leakage}.  The theory results also show that there is  a range of parameters for which a reduced IR perturbation leads to larger RPP peaks. In the calculations, this happens because the IR-controlled line shape corresponding to the two different perturbations is different. In particular, the larger perturbation leads to a line shape which has a deeper minimum on line center at the propagation distance when the RPP peak starts to appear. This means that the RPP peak has to grow from a lower OD value than in the smaller perturbation case, in which the line shape starts out more asymmetric and with a shallower minimum. This difference in the basic shape can still be recognized at the end of the medium as in Fig.~\ref{FigPolarization} even though the OD line shape has undergone substantial change during the propagation. This difference in the line shapes can also be seen in the experimental results, where the basic line shape in the perpendicular case is more dispersive (indicating a phase shift of approximately $\pi/2$), whereas the line shape in the parallel case has a more pronounced peak on the high-energy side, indicating a phase shift larger than $\pi/2$ (the limiting, symmetric, case of a window-resonance-like shape would correspond to a phase shift of $\pi$). These polarization dependence results thus also indicate that for dense gases, the RPP effect can have a substantial effect on how the IR perturbation manifests itself in the absorption spectrum and should be taken into account if the absorption line shape is to be used to characterize IR-induced dynamics.  

%

\section{Summary}
\label{Summary}

In this paper, we have presented a joint experimental and theoretical study of transient absorption of an APT by a dense, helium gas. We have systematically explored the evolution and modification of the absorption line shape with IR-XUV delay, IR intensity, gas pressure, and the relative polarization of the  XUV and IR pulses. We have focused on elucidating how the interplay between the microscopic IR-laser-dressing and the macroscopic RPP reshaping affects the transient absorption of the attosecond XUV pulse. To this end we have employed both time- and frequency-domain descriptions of the light-matter interaction. We have demonstrated that the absorption line shape is controlled not only by the IR pulse through its delay and intensity as it is in the single atom limit, but is also influenced by collective effects induced during propagation through the resonant medium. For small optical thicknesses, we have shown (as in \cite{Liao2015}) that RPP leads to the appearance of narrow absorption features on line center which are absent in the single atom response. For larger optical thicknesses, RPP effects can lead to substantial dynamical reshaping of the absorption line shape, in particular when the duration of the IR pulse is long compared to the duration of the first sub-pulse in the propagating XUV light. The reshaping can be understood in terms of an effective reduction of the LIP, since a long IR pulse means that only part of the total LIP will be induced over the first XUV sub-pulse. We also showed that the presence of the dressing IR field influences the resonant XUV propagation dynamics. In particular, we showed that for long IR pulses, the presence of the dressing field changes the time structure of the propagating XUV pulse. 

The results in this study demonstrate that the full description of the interaction of an attosecond XUV pulse with a dressed, optically thick medium must include a truly dynamical feedback between the dipole moment of the individual atoms and the time structure of the propagating pulse. Our study has illustrated and explained what we believe to be the basics of how the XUV transient absorption spectrum develops as a function of XUV-IR delay, IR intensity, and gas pressure. This understanding of the fundamental but complicated features of transient photo absorption will be crucial in extending this powerful time-resolved spectroscopic tool to more complicated and high-density systems such as plasmas and condensed phase systems.  

\begin{acknowledgments}
This work was supported by the U. S. Army Research Laboratory and the U. S. Army Research Office under grant number W911NF-14-1-0383, and by the U. S. Department of Energy, Office of Science, Office of  Basic Energy Sciences, under Award No. DE-FG02-13ER16403. C.T.L. acknowledges
support from Arizona TRIF Imaging Fellowship. Computing resources were provided by the High Performance Computing Centre at LSU.
\end{acknowledgments}

\bibliography{Liao_Helium2_Refs}


\end{document}